\documentclass[sigconf]{acmart}
\AtBeginDocument{%
  \providecommand\BibTeX{{%
    \normalfont B\kern-0.5em{\scshape i\kern-0.25em b}\kern-0.8em\TeX}}}

\copyrightyear{2021}
\acmYear{2021}
\setcopyright{acmcopyright}\acmConference[Accepted at SSDBM 2021]{33rd International Conference on Scientific and Statistical Database Management}{July 6--7, 2021}{Tampa, FL, USA}
\acmBooktitle{33rd International Conference on Scientific and Statistical Database Management (SSDBM 2021), July 6--7, 2021, Tampa, FL, USA}
\acmPrice{15.00}
\acmDOI{10.1145/3468791.3468810}
\acmISBN{978-1-4503-8413-1/21/07}

\usepackage{multirow}

\usepackage[utf8]{inputenc}
\usepackage[linesnumbered,ruled,vlined]{algorithm2e}

\usepackage{todonotes}

\usepackage{graphicx}

\begin{document}

\title{MoParkeR : Multi-objective Parking Recommendation}


\author{Mohammad Saiedur Rahaman}
\affiliation{%
  \institution{RMIT University}
  \streetaddress{GPO Box 2476}
  \city{Melbourne}
  \state{VIC}
  \country{Australia}
  \postcode{3000}
}\email{saiedur.rahaman@rmit.edu.au}
\author{Wei Shao}
\affiliation{%
  \institution{RMIT University}
  \streetaddress{GPO Box 2476}
  \city{Melbourne}
  \state{VIC}
  \country{Australia}
  \postcode{3000}
}\email{wei.shao@rmit.edu.au}

\author{Flora D. Salim}
\affiliation{%
  \institution{RMIT University}
  \streetaddress{GPO Box 2476}
  \city{Melbourne}
  \state{VIC}
  \postcode{3000}
  \country{Australia}
}\email{flora.salim@rmit.edu.au}

\author{Ayad Turky}
\affiliation{%
  \institution{RMIT University}
  \streetaddress{GPO Box 2476}
  \city{Melbourne}
  \state{VIC}
  \country{Australia}
  \postcode{3000}
}\email{ayad.turky@rmit.edu.au}
\author{Andy Song}
\affiliation{%
  \institution{RMIT University}
  \streetaddress{GPO Box 2476}
  \city{Melbourne}
  \state{VIC}
  \country{Australia}
  \postcode{3000}
}\email{andy.song@rmit.edu.au}

\author{Jeffrey Chan}
\affiliation{%
  \institution{RMIT University}
  \streetaddress{GPO Box 2476}
  \city{Melbourne}
  \state{VIC}
  \country{Australia}
  \postcode{3000}
}\email{jeffrey.chan@rmit.edu.au}

\author{Junliang Jiang}
\affiliation{%
  \institution{RMIT University}
  \streetaddress{GPO Box 2476}
  \city{Melbourne}
  \state{VIC}
  \country{Australia}
  \postcode{3000}
}\email{junliang.jiang@rmit.edu.au}
\author{Doug Bradbrook}
\affiliation{%
  \institution{Mornington Peninsula Shire Council}
  \city{Rosebud}
  \state{VIC}
  \country{Australia}
  \postcode{3939}
}\email{doug.bradbrook@mornpen.vic.gov.au}

\renewcommand{\shortauthors}{Rahaman, et al.}

\begin{abstract}

Existing parking recommendation solutions mainly focus on finding and suggesting parking spaces based on the unoccupied options only. However, there are other factors associated with parking spaces that can influence someone's choice of parking such as fare, parking rule, walking distance to destination, travel time, likelihood to be unoccupied at a given time. More importantly, these factors may change over time and conflict with each other which makes the recommendations produced by current parking recommender systems ineffective. In this paper, we propose a novel problem called multi-objective parking recommendation. We present a solution by designing a multi-objective parking recommendation engine called \emph{MoParkeR} that considers various conflicting factors together. Specifically, we utilise a non-dominated sorting technique to calculate a set of Pareto-optimal solutions, consisting of recommended trade-off parking spots. We conduct extensive experiments using two real-world datasets to show the applicability of our multi-objective recommendation methodology.
  
\end{abstract}




\maketitle
\section{Introduction}
Finding street parking in crowded cities is challenging and thus requires a highly functional and efficient parking management strategy. Searching for available parking can cause congestion and wasted land use. A research found that vehicles cruising for suitable parking in Los Angles were responsible for 730 tons of carbon emissions through the burning of 47,000
gallons of gasoline in one year \cite{mathur2010parknet}. The research also identified that the distance travelled during the search for parking was equivalent to 38 trips around the world. These problems can be minimised by providing drivers with effective recommendations about their parking choices. However, effective parking recommendation is very complex as it is influenced by various factors such as parking fare, rule, walking distance to the destination location, total travel time, and likelihood of parking locations be available at a given timestamp. Moreover, these factors can be conflicting with each other. For instance, a low-fare parking spot may require long-distance walking to destination. Additionally, some of these factors (e.g. fare, parking rules) may vary across different times of the day which makes the parking recommendation problem more complex.
%\cite{rajabioun2015street, shao2018parking, yang2019deep, salim2015urban}
With the growing trends in adopting ubiquitous technologies, many cities around the world have implanted smart sensors to collect data related to parking events. Many researchers use this data to extract meaningful information. The current research mainly focuses on predictive analytics of parking events using univariate and multi-variate signals. Another research direction uses event logs from parking sensor data to make new policies (e.g. new parking rule or dynamic parking price)~\cite{ qian2014optimal, lei2017dynamic}. There are a few research that provides parking recommendation based on future availability \cite{vlahogianni2016real}. However, this group of research rarely consider user preferences as it is challenging and complex due to the presence of multiple factors. The presence of various factors in user preferences makes the existing parking recommender systems ineffective.

%inci2015review,inci2018parking,zheng2015parking,

In this paper, we address the challenges associated with parking recommendations by defining a multi-objective optimization problem. We develop a recommendation engine that takes users' conflicting preferences as input and provides a set of recommended parking spots in the Pareto-front. We leverage two real-world large parking datasets to extract conflicting factors and compute personalized parking recommendations. In particular, the contributions of this paper are as follows:

\begin{itemize}
\item Introduce a new parking recommendation problem called multi-objective parking recommendation which has a range of prospective implications for smart cities.
\item Design of a multi-objective parking recommendation engine based on a non-dominated sorting technique. This approach is further refined by adopting a crowding distance mechanism with objective-thresholding.
% \item A spatial clustering of parking sensors based on network distances to identify parking lots and computation of a set of dynamic features associated with each parking lot.
\item Design a system prototype that is able to compute and present end-to-end travel routes connecting recommended parking spots.
\end{itemize}

%The rest of this paper is organized as follows. The related literature including parking availability prediction, parking dynamics, parking policy, and multi-objective recommendation are discussed in Section 2. We present \textit{MoParkeR}, a novel multi-objective parking recommendation engine in Section 3. Section 4 discusses a series of real-world deployments and experiments which is followed by a detailed discussion on findings in Section 5. Finally, the paper concludes with a direction to future works in Section 6.

%\paragraph{Functionality} The Elsevier article 

\section{Related Literature}
This section discusses current research utilizing parking sensor data and the application of multi-objection recommendations in various fields. We also highlight the challenges of integrating multi-objective optimization in parking recommendations considering the current state of research.

%\subsection{Parking availability prediction}
%\cite{lin2017survey, shi2017study, vlahogianni2016real, shao2020incorporating, wu2020iot}.

Finding an available car park is challenging for urban drivers. To help drivers find parking locations in the CBD areas, various research has been conducted that use probabilistic and machine learning models to predict parking occupancy or duration. In this cohort of research, various types of sensor feed (e.g. vision-based sensors, physical sensors) are used Since most parking data are recorded from in-ground sensors, many researchers tend to study on extracting features from time-series data and spatio-temporal data to explore the correlation between contextual information and parking availability \cite{shao2016clustering,ahmed2019blockchain, rahaman2018wait}. %Parking availability are influenced by many factors such as traffic congestion \cite{inci2017external, rahaman2018wait}, fares \cite{yan2018truthful}, and nearby Point of Interests.  %Nevertheless, most parking systems in association with many other features such as PoI, parking duration, the penalty for violation can provide opportunities for researchers to explore the correlation between parking availability and these contexts. 

%\subsection{Parking policy} Parking management plays a significant role in the parking recommendation area. Researchers in this area purpose many model-based methods to manage and monitor the parking situation in the past \cite{xiao2018restrain}. Some researchers focus on the price and economic impact on parking policy \cite{ verhoef1995economics, small1997economics}. Aside from the traditional methods, data-driven methods have become increasingly popular due to their effectiveness, employed in many studies, such as a study on parking policy based on the transportation data and other contextual data \cite{xu2017implementation}. 

%inci2018parking, inci2015review,
%\subsection{Multi-objective recommendation}

Recommender systems suggest items so that the users' potential interests are met or at least optimized. In multi-objective recommendation, users have more than one preference to be considered by the recommendation engine. Usually multi-objective recommendation is considered as a multi-objective optimization problem \cite{deb2001multi}. One of the approaches to address this problem is to find a set of Pareto-optimal solutions. Several domain-specific research has been conducted in this area \cite{rahaman2017capra,turky2016multi}.

\section{MoParkeR System Overview}

\subsection{Datasets}

To investigate the multi-objective parking recommendation, this research utilises smart parking datasets from two city councils in Victoria, Australia. The first dataset is a continuous log of more than 4650 in-ground parking sensors around the Melbourne CBD area. The sensors are capable of detecting parking events. A total of 5.9 million records are present in the 2017-18 parking dataset which is publicly available through the open data portal of the City of Melbourne\footnote{https://data.melbourne.vic.gov.au/Transport/On-street-Car-Parking-Sensor-Data-2017/u9sa-j86i}.

\iffalse
\begin{table}[h]
\small
\centering
\renewcommand{\arraystretch}{0.7}
\caption{Description of attributes in parking datasets.}
\label{tab:datasets}
\begin{tabular}{ll}
\toprule
\textbf{Features} & \textbf{Description}\\\midrule \midrule
Street Marker &  Unique id for the parking bay\\ 
%City Area & Administrative region a parking bay belong to \\      
 
Longitude & Geographical information \\ 
Latitude & Geographical information      \\
Arrival time & Start time of a parking event \\ 
Departure time & End time of a parking event \\ 
Duration  & Duration of a parking event \\\bottomrule                             
\end{tabular}
\end{table}
\fi

The second dataset was collected from the city of Rye as part of Victorian government's Smart City and Smart Suburb project. This dataset integrates parking-event feeds from in-ground as well as vision-based sensors. Both of the datasets have similar attributes to describe a particular parking event in a time-series manner. %The relevant common fields used for our multi-objective parking recommendation are listed in Table \ref{tab:datasets}.

\begin{figure}[h]
  \centering
  \includegraphics[width=\columnwidth]{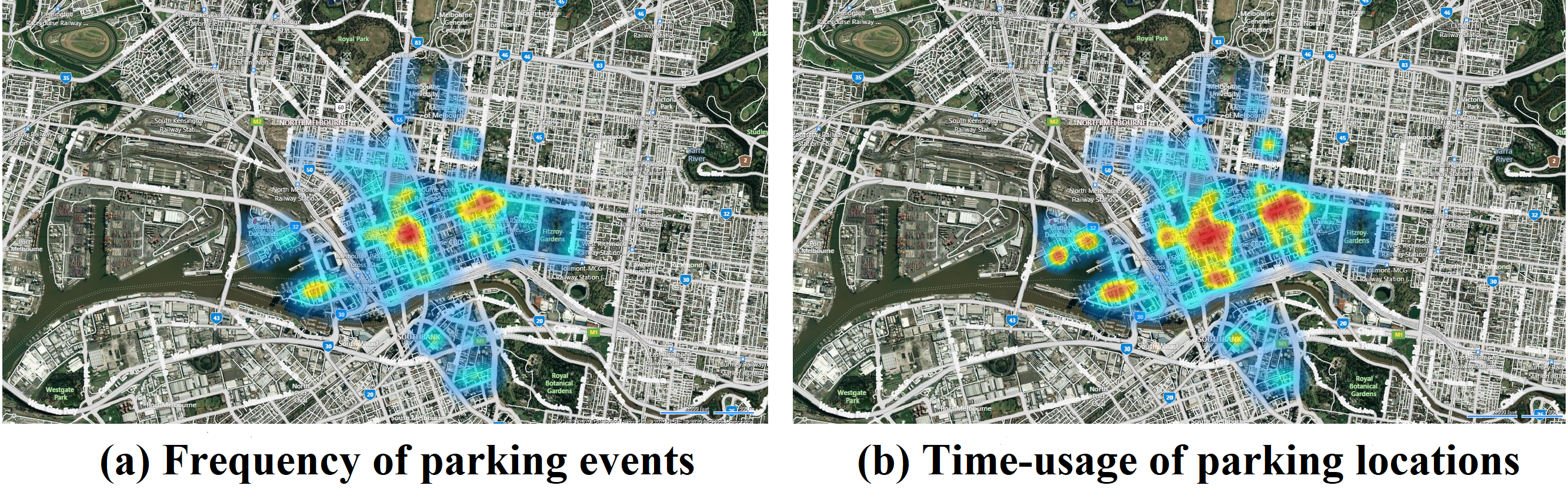}
  \caption{Heatmap showing (a) frequency of parking events and (b) time-usage of parking locations between 8.30 am and 9.00 am in Melbourne CBD.}
  \label{fig:parking_mel}
\end{figure}

\begin{figure}[h]
  \centering
  \includegraphics[width=\columnwidth]{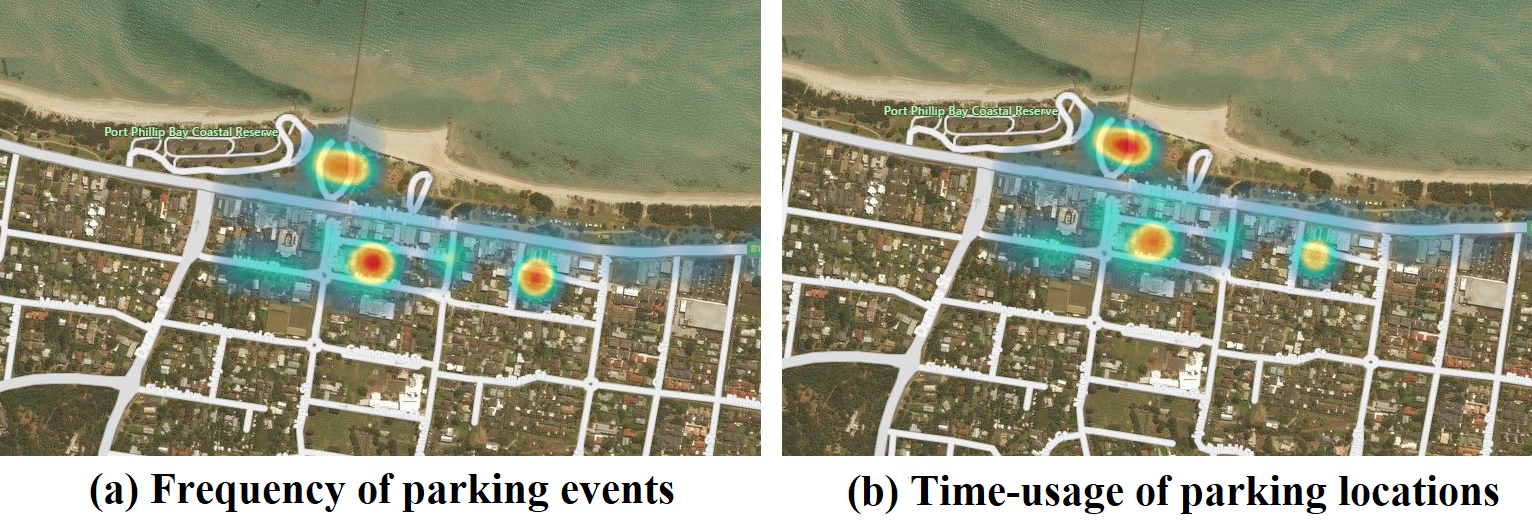}
  \caption{Heatmap showing (a) frequency of parking events and (b) time-usage of parking locations between 02.00 pm and 02.30 pm in Rye.}
  \label{fig:parking_mornington}
\end{figure}

To understand the parking behavior in these two cities, we plot heatmaps of parking locations in terms of frequencies of parking events and time-usage. Figures \ref{fig:parking_mel} and \ref{fig:parking_mornington} illustrate different parking patterns in these two cities. The heatmap in Figure \ref{fig:parking_mel}(a) shows average parking events of two-weeks between 8.30 am and 9.00 am in different parts of Melbourne CBD while Figure \ref{fig:parking_mel}(b) depicts a time-usage heatmap of parking locations. We can see that the parking locations with a high number of parking events are also accounted for high utilisation times. However, there are many other locations with lower parking events that turned out to be highly utilised by single parking events within the selected time window. In contrast, similar locations seemed to be highly utilised in terms of both events and time-usage in Rye as can be seen from Figures \ref{fig:parking_mornington}(a)-(b). We plot the average utilisation in the first two weeks of January 2020 between 2.00 pm and 2.30 pm. This may be due to the fact that Rye is a busy tourist spot on the coastline where these parking locations provide easy access to the beaches and shops.

%%%%%%%%%%%%%%%%%%%%%%%%%%%%%%%%%%%%%%%%%%%%%%%%%%%%%%
%%%%%%%%%%%%%%%%%%%%%%%%%%%%%%%%%%%%%%%%%%%%%%%%%%%%%%
\iffalse
\begin{figure}[h]
  \centering
  \includegraphics[width=0.9\textwidth]{./figures/parking_slots}
  \caption{Parking locations in Melbourne CBD.}
  \label{fig:parking_slots}
\end{figure}
%%%%%%%%%%%%%%%%%%%%%%%%%%%%%%%%%%%%%%%%%%%%%%%%%%%%%%
\begin{figure}[h]
  \centering
  \includegraphics[width=0.9\textwidth]{./figures/parking_events}
  \caption{Distribution of parking locations with a varying number of parking events.}
  \label{fig:parking_slots}
\end{figure}
%%%%%%%%%%%%%%%%%%%%%%%%%%%%%%%%%%%%%%%%%%%%%%%%%%%%%%
\begin{figure}[h]
  \centering
  \includegraphics[width=0.9\textwidth]{./figures/average_occupied_time}
  \caption{Distribution of average occupied times in different parking locations between 8.30 am to 9.30 am.}
  \label{fig:parking_slots}
\end{figure}
\fi
%%%%%%%%%%%%%%%%%%%%%%%%%%%%%%%%%%%%%%%%%%%%%%%%%%%%%%
%%%%%%%%%%%%%%%%%%%%%%%%%%%%%%%%%%%%%%%%%%%%%%%%%%%%%%

\subsection{Multi-objective parking recommendation}

\begin{figure}[h]
  \centering
  \includegraphics[width=\columnwidth]{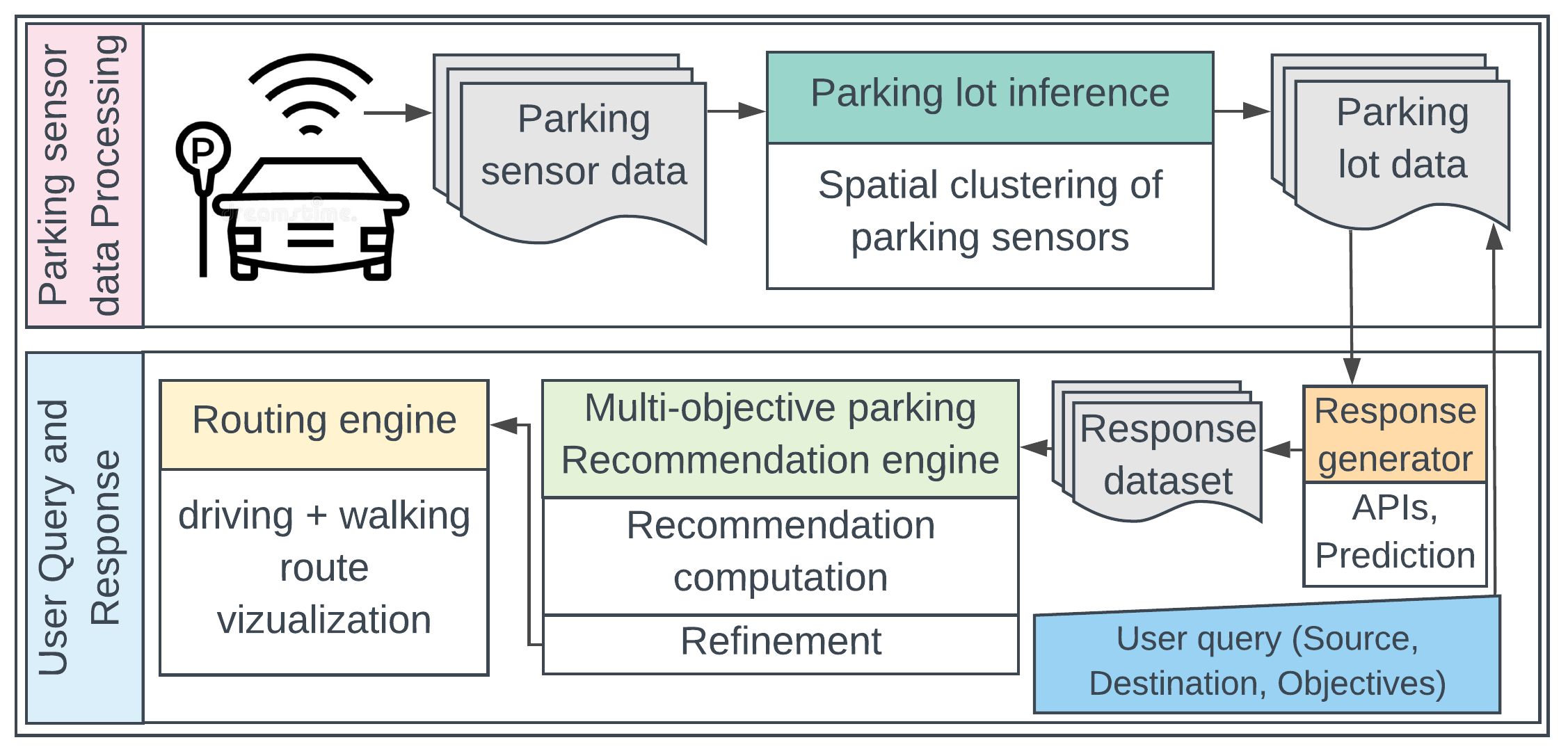}
  \caption{An overarching view of the \emph{MoParkeR} recommendation engine.}
  \label{fig:MOPR}
\end{figure}

In this section, we present \emph{MoParkeR}, a multi-objective parking recommendation engine. Our parking recommendation engine has two main components: i) parking sensor data processing and ii) user query and response. The key components of the \emph{MoParkeR} is given in Figure \ref{fig:MOPR}.

%\sr{add response generator}

\subsubsection{Parking sensor data processing} In this module, parking events data collected from sensors (i.e. in-ground and vision-based) are stored in the central storage. Since two side-by-side parking bays are of less significance in terms of their physical location, we cluster those to form parking lots. We adopted a spatial-clustering technique to form parking lots consists of multiple parking bays. Specifically, two parking bays are part of a single parking bays (i.e. cluster) if both are connected to each other, or the distance between these slots is within a specific value and they have the same parking restriction (e.g. 1 hour, 2 hours, loading zone). This processed data is then stored in a parking lot database which is used to handle user queries.

\subsubsection{User query and response} This module is responsible for processing user queries to provide parking recommendations. As can be seen from Figure \ref{fig:MOPR}, a user-defined query consists of a user's source location, the destination location, and a set of objectives. Once a query is defined, it is sent to the parking sensor data processing module which returns the locations of parking lots. The APIs in response generator utilises these locations of the parking lots to compute the objective values and create a response dataset. In practice, the likelihood of parking spaces needs to be estimated from real-time data. One of the APIs plays as a sub-module of the `Response generator' which is responsible for Likelihood prediction of parking spaces. This prediction API uses state-of-the-art techniques to generate likelihood prediction.

To provide a multi-objective parking recommendation, this research computes four response factors associated with each parking lot. First, we consider the total travel time from source to destination for a recommended parking lot. Given a source location $v_i$, destination location $v_k$, and any parking lot $v_j$, the total travel time is defined as follows.

\begin{equation}
    T(j) = t(v_i, v_j) + t(v_j,v_k); t(.) \geq 0
\end{equation}

Here, $T(j)$ is the total travel time if a drive needs to park at parking lot $v_j$. $t(v_i, v_j)$ is the travel time from source location $v_i$ to parking lot $v_j$ while $t(v_j,v_k)$ denotes travel time between parking lot $v_j$ and destination $v_k$.

Second, we define the required walking distance from any parking lot $v_j$ to destination $v_k$ as follows.

\begin{equation}
    W(j) = d(v_j,v_k); d(.) \geq 0
\end{equation}

Here, $d(v_j,v_k)$ is the walking distance between $v_j$ and $v_k$.

Third, for a parking duration of $s$ in a parking lot $v_j$, we consider the dynamic parking fare as follows.

\begin{equation}
    F(j) = f_s
\end{equation}

Here, $f_s \geq 0$ is the parking fare for $s$ duration of parking at $v_j$.

Fourth, we compute the likelihood of a parking spot $j$ being available as follows.

\begin{equation}
    L(j) = 1- \frac{\sum_{i=1}^{n} O_{t}(i)}{n*\tau}
\end{equation}

Here, $O_{t}(i) \geq 0$ and $0 \leq L(j) \leq 1$; $O_{t}(i)$ is the occupied time of any parking bay $i$ within a parking lot $j$, $n$ is the number of parking bays within $j$, and $\tau$ is the length of the time window when a driven will arrive at $j$. Therefore, $\frac{\sum_{i=1}^{n} O_{t}(i)}{n}$ is the average occupied time of parking lot $j$.

Once the responses are computed using Eq. 1-4, it is forwarded to the multi-objective parking recommendation engine which computes recommended parking lots. The recommended lots are used by the routing engine to provide routing support to the user.\\

\noindent\textbf{Multi-objective parking recommendation engine}. A driver who is planning to drive from a source to a destination needs to find a suitable parking location to park her car before walking to the destination. The suitability of a parking location can be dependent on various crucial factors including the total travel time, required walking distance from any parking lot to the destination, parking fare, and likelihood of getting a parking lot available. We assume the user would prefer the path that connects a parking lot with lower parking fare, shorter travel time and walking distance, and a higher likelihood of getting an available bay in a parking lot. However, in practice, these factors may be in conflict with each other. In this case, one should provide a set of trade-off parking lots, which are termed the \emph{pareto-optimal parking lots}, instead of one single global optimal solution.
 
In our research, we have three objectives to minimize and one objective to maximize. The four objectives to be minimized or maximized in parking recommendation can be described as follows:
    
    \begin{eqnarray}
    \min_j T(j) = t(v_i, v_j) + t(v_j,v_k),
        \label{eq:f1} \\
    \min_j W(j) = d(v_j,v_k),
        \label{eq:f2} \\
        \min_j F(j) = f_s,
        \label{eq:f3} \\
    \max_j L(j) = 1- \frac{\sum_{i=1}^{n} O_{t}(i)}{n*\tau}
        \label{eq:f4} 
    \end{eqnarray}
    
where, $T(j)$ is the total travel time from source $v_i$ to destination $v_k$ if parking lot $v_j$ is chosen, $W(j)$ is the walking distance from $v_j$ to destination, and $F(j)$ is the fare of $v_j$, and $L(j)$is the likelihood of getting an free parking bay at $v_j$.

Given two parking lots $v_{j,1}$ and $v_{j,2}$, $v_{j,1}$ is said to \emph{dominate} $v_{j,2}$ if and only if all the objective values of $v_{j,1}$ are no worse than those of $v_{j,2}$, and there is at least one objective for which $v_{j,1}$ has a better value than $v_{j,2}$. We denote $v_{j,1} \prec v_{j,2}$ for $v_{j,1}$ dominating $v_{j,2}$. A parking lot $V_{j^*}$ is said to be \emph{Pareto-optimal}, if and only if there is no other parking lot that dominates $V_{j^*}$. The goal of this problem is to find all the possible Pareto-optimal parking lot and provide a smaller list as recommendation.

%pareto:matthewj

In this paper, we employed an epsilon-nondomination based multi-objective sorting algorithm to find the Pareto-optimal parking lots. The details of epsilon-domination framework are described in \cite{laumanns2002combining,10.1162/106365605774666895}. Since there could be many elements in the Pareto front, we adopted a \textit{Crowding distance} technique to achieve a smaller number of parking lots for final recommendation. The concept of crowding distance is to measure the relative isolation of a solution in the Pareto front from other solutions. The greater the crowding distance, the greater is its isolation and hence, higher chance to be included in the final recommendation. We further refine the selected solutions by applying an objective-thresholding. Specifically, we consider a threshold for an objective, e.g., for parking likelihood in our case to ensure that the solution subset only comprises of the parking spaces with high likelihood. Algorithm \ref{algo:crowdingD} illustrates the deployed crowding distance calculation mechanism in our study. Given a set of candidate solutions $S$ and objective-threshold $\tau_{m}$, this algorithm computes crowding distance of all candidate solutions in $S$. The computation starts by initializing all candidates to $0$. Upon satisfying the objective-threshold $\tau_{m}$, the candidates are sorted in ascending order in turn for each objective. Note that the first and last candidates (i.e. boundary points) from the sorted list are selected automatically as they have an only neighbor candidate in total. Hence, we assign the largest crowding distance score of $\infty$. Give any objective, the crowding distance for the rest of the candidates $i$ in $S$ are calculated as follows. First, we subtract the objective value of candidate $i-1$ from the objective value of candidate $i+1$. Then the result is divided by the ${Max(S.m)-Min(S.m)}$ where, $Max(S.m)$ and $Min(S.m)$ are the maximum and minimum objective values in $S$ for objective $m$. The final crowding distance is computed by repeating this process and taking the sum over all of the objectives.  

\begin{algorithm}[]
\footnotesize
\KwIn{$S$: set of all solutions, $\tau_{m}$: threshold for objective $m$}
\KwOut{$CD_{s[i]}$ \tcp*{crowding distance for all $i$ in $S$}}
  
 int $l=|S|$ \tcp*{$l$ is the number of solutions in $S$}
 $\tau_{m}.initialize()$ \tcp*{initialize $\tau_{m}$}
\ForEach{$i \in S$}    
    { 
    	$CD_{s[i]}= 0$ \tcp*{initialize crowding distance}
    }

\If {$m$ satisfies $\tau_{m}$}{
\ForEach{$m$} 
    {
       S = $sort(S,m)$\;
    	$CD_{s[1]}$, $CD_{s[l]}$ = $\infty$ \tcp*{select boundary points}
    	
    	\For{$i = 2$ to $l-1$}    
            { \tcc{for all other solutions in $S$}
    	        $CD_{s[i]} = CD_{s[i]} + \frac {{s[i+1]}.m - {s[i-1]}.m}{Max(S.m)-Min(S.m)}$ \;
    	    }

    }
 return $CD_{s[i]}$\; 
}
 \caption{Crowding distance calculation}
 \label{algo:crowdingD}
\end{algorithm}

    \noindent\textbf{Routing engine.} Once computed, the recommended parking lots are considered to generate journeys from any source location to the destination location connecting all recommended parking lots. Specifically, driving routes from the source location to parking lots are constructed before generating walking routes from parking lots to the destination location. For route generation, APIs such as Google Map can be utilised.

%Data preprocessing and query-based adaptation:
%Objective evaluation:
%Parking recommendation:

%https://link.springer.com/chapter/10.1007/978-3-540-70928-2_60
%https://arxiv.org/pdf/1910.08751.pdf

\section{Experimental Studies}

In this section, we present the experiential results of the proposed approach. To evaluate our approach, we develop a system prototype and conducted two case studies for two cities in Victoria, Australia including the city of Melbourne and the city of Rye. These cities are good examples for the experimental studies as one of these is a busy business district while the other is a popular tourist spot. %In addition, different city layouts such as urban and suburban are taken into account. For our multi-objective parking recommendation experiments, we selected two random journeys that include a parking spot between source and destination locations. 

%Note that there is no existing system that considers a parking spot in the middle of a journey that satisfies various conflicting preferences of users. In addition, since we have developed \emph{MoParkeR} that employs epsilon-domination based sort, it is guaranteed to find all the trade-off parking lots. We also employ a crowding distance mechanism to refine the computed Pareto front of parking lots to present a smaller number of recommendations. The effectiveness of the recommendation can be further improved by adopting a threshold-based approach for each objective. For instance, we considered parking lots with an availability likelihood of 0.7 or higher in our experiments. This is to increase the chance of getting a parking space available within the recommended set of trade-off parking spaces.

\subsection{Case Study-1: Melbourne, VIC, Australia}

\begin{figure}[h]
  \centering
  \includegraphics[width=\columnwidth]{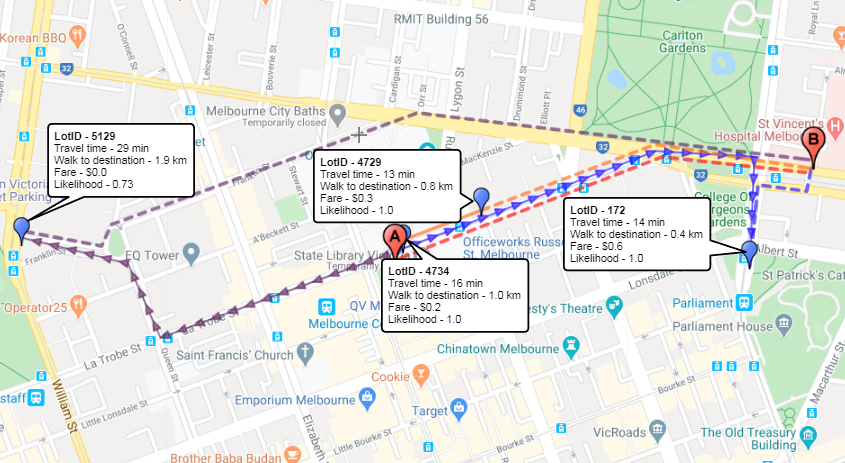}
  \caption{Routes with four recommended parking lots in Melbourne CBD, VIC, Australia [source location: State Library Victoria (A); destination location: St Vincent's hospital (B)] }
  \label{fig:melbourne-stateLib2stVincentH}
\end{figure}

%10.45, 8th February 2017
In this case study, we find connecting parking lots for journeys between the source location, State Library (point A), and destination location, St Vincent's hospital (point B). These two locations are situated in inner-city Melbourne. Figure \ref{fig:melbourne-stateLib2stVincentH} shows four journeys connecting four different parking lots computed by our \emph{MoParkeR} engine. These parking lots are chosen arbitrarily from the set of recommended parking lots for illustration purposes. A full list of recommended trade-off parking lots are given by Table \ref{tab:pareto:exp-1}.

\begin{table}[h]
\centering
\renewcommand{\arraystretch}{0.7}
\caption{Recommended parking Lot IDs in the Pareto front (City of Melbourne-1).}
\label{tab:pareto:exp-1}
%\resizebox{\linewidth}{!}{%
%\scalebox{0.8}{
\resizebox{0.95\linewidth}{!}{%
\begin{tabular}{ccccc}
\toprule
\textbf{LotId} & \textbf{Travel Time (min)} & \textbf{Walk Dist. (km)} & \textbf{Fare (\$)} & \textbf{Likelihood (\%)} \\\midrule \midrule
172  &         14     &         0.4 &  0.6      &            1.0  \\ 
4729   &        13      &        0.8  & 0.3       &           1.0  \\ 
4734    &      16       &       1.0   & 0.2       &           1.0  \\ 
5129     &     29        &      1.9    & 0.0       &           0.73 \\
4716  &    35  &     2.1  & 0.0  &  0.92 \\\bottomrule

\end{tabular}
}
\end{table}

% CoM1 -12 (5)
% CoM2 -17 (5)

%14.00, 8th February 2017

To generate these journeys, we first applied \emph{MoParkeR} to calculate a set of non-dominated (i.e., Pareto-optimal) parking lots. The computed Pareto front can have a large number of parking lots since we have four objectives to consider. To reduce the users' information burden, this number is further reduced by applying a crowding distance mechanism and objective-threshold in the computed Pareto-front before presenting the final recommendations. Upon calculation of final recommendations, the journeys can be constructed by using direction services (e.g. Google API). We can see from Figure \ref{fig:melbourne-stateLib2stVincentH} that the four recommended parking LotIDs have non-dominated objective values in terms of travel time, walk to a destination, parking fare, and parking likelihood. For instance, the LotID-172 requires a driver smallest walking to the destination (i.e. 0.4 km) with an end-to-end travel time of 14 minutes, a fare of \$0.6 and a very certain likelihood of parking. On the other hand, LotID-5129 provides users with a free parking lot but costs a long walk to destination after parking the car which also causes a longer travel time. All these trade-off parking lots satisfy our recommendation refinement criteria (i.e. likelihood $\geq$ 0.7) for an increased chance of getting a parking spot available to park.

\subsection{Case Study-2: Rye, VIC, Australia}

\begin{figure}[h]
  \centering
  \includegraphics[width=\columnwidth]{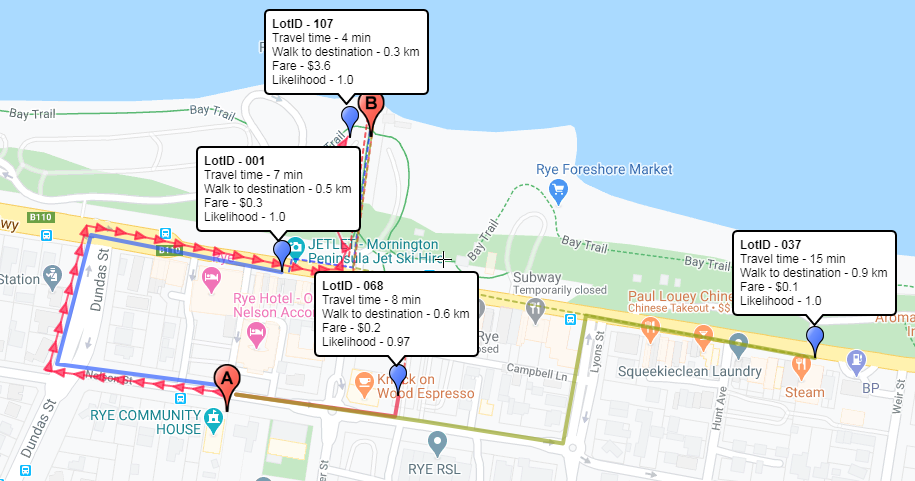}
  \caption{Routes with four recommended parking lots in Rye, VIC, Australia [source location: Rye Community House, Nelson Street, Rye VIC (A); destination location: Rye Pier, Bay Trail, VIC (B)] }
  \label{fig:mornington-map}
\end{figure}

This case study was conducted by selecting a random pair of source and destination locations in Rye which is a tourist hot-spot of regional Victoria in Australia. We employ \textit{MoParkeR} recommendation engine to construct journeys from Rye community house to Rye pier on a Saturday afternoon (i.e., 11th of January 2020) as it was peak-time for tourist visits before COVID-19 lock-down in Australia. These locations are marked in Figure \ref{fig:mornington-map} as A and B respectively. \textit{MoParkeR} computed seven trade-off parking lots to construct journeys between A and B. Our \emph{MoParkeR} engine constructed seven journeys from A to B connecting all of these parking lots. For illustration purpose, four journeys are projected in Figure \ref{fig:mornington-map}. We can see that all these recommended parking lots reside within four different routes are non-dominating to each other in terms of our defined conflicting objectives (i.e. minimal travel time, minimal walk to destination, minimal parking fare and maximal parking likelihood) and all of those satisfy the objective-threshold for likelihood which is 0.7 or more in our experiments.

\begin{table}[h]
\centering
\renewcommand{\arraystretch}{0.7}
\caption{Recommended parking Lot IDs in the Pareto front (Rye).}
\label{tab:pareto:mornington}
%\resizebox{\linewidth}{!}{%
%\scalebox{0.8}{
\resizebox{0.95\linewidth}{!}{%
\begin{tabular}{ccccc}
\toprule
\textbf{LotId} & \textbf{Travel Time (min)} & \textbf{Walk Dis. (km)} & \textbf{Fare (\$)} & \textbf{Likelihood (\%)} \\\midrule \midrule
001  &     7     &         0.5 &  0.3      &            1.0  \\
037   &      15      &        0.9  & 0.1       &           1.0  \\
068    &   8       &       0.6   & 0.2       &           0.97  \\
107     &     4        &      0.3    & 3.6      &           1.0 \\
003     &    6        &      0.5   & 2.3 &                 1.0  \\
109   &      7      &        0.3  &  0.8 &                  1.0 \\
110  &      7      &        0.4   & 0.5  &               1.0\\\bottomrule
\end{tabular}
}
\end{table}

\subsection{Comparison with Baseline}

To demonstrate the effectiveness of \textit{MoParkeR}, we conduct a utility analysis. First, we conduct a utility comparison with currently available parking recommendation solutions. As can be seen from Table \ref{tab:comparison}, no other solutions except \textit{MoParkeR} provides trade-off parking lots considering parking cost, walking distance to destination, total travel time from the source location to destination and likelihood of getting a parking lot available. One of the solutions considers two utilities (i.e. parking cost and proximity to destination). However, since this approach does not consider parking likelihood, the recommendation can be ineffective when the driver arrives at the parking location with no availability.

\begin{table}[t]
\centering
\caption{Utility comparison with parking services}
\label{tab:comparison}
\resizebox{0.95\linewidth}{!}{%
\begin{tabular}{lccccc}
\toprule
\multirow{2}{*}{\textbf{\begin{tabular}[c]{@{}l@{}}Parking\\ service\end{tabular}}} & \multicolumn{4}{c}{\textbf{Utilities}}                                                                         \\ \cmidrule{2-5} 
    & \textbf{Parking cost} & \textbf{Walk to dest.} & \textbf{Likelihood} & \textbf{Travel time}  \\ \midrule \midrule
iParker \cite{kotb2016iparker}  & $\checkmark$  &     &        &            \\ 
Smart Parking \cite{geng2013new} &  $\checkmark$  & proximity to dest.    &      &          \\ 
Parking Assignment \cite{kim2019parking}     &  $\checkmark$ &     &      &                 \\ 
Parking rank \cite{dong2018parking}   & $\checkmark$ & &  parking space count &            \\ 
ROSAP \cite{mejri2016reservation}   & &  $\checkmark$ &    &  \\ 
Park Indicator \cite{parikh2018park}               &     proximity to parking &  &   &       \\ 
\textbf{MoParkeR}  & $\checkmark$ &$\checkmark$  & $\checkmark$& $\checkmark$ \\ \bottomrule
\end{tabular}
}
\end{table}

Additionally, since there is no standard approach for multi-objective parking recommendation, we design a greedy approach to conduct a further comparison with \textit{MoParkeR}. Unlike considering the proximity of the parking lots in relation to the current location of the vehicle, the greedy approach recommends a parking slot based on a predefined proximity threshold with respect to the destination location. 

\iffalse
Algorithm \ref{algo:greedy} illustrates the designed greedy approach. Given a destination location $D$, proximity threshold to destination $p_x$, and a set of parking lots $L$, Algorithm \ref{algo:greedy}, randomly select and recommend a parking lot $L_r$ $\in L$.

\begin{algorithm}[h]
\footnotesize
\KwIn{Destination $D$, proximity threshold $p_x$, set of parking lots $L$}
\KwOut{A recommended parking lot $L_r$ $\in L$}
  
 $S = \emptyset$  \tcp*{ initialize $S$, the set of all solutions}
  
\ForEach{$j \in L$}    
    { 
    	$Dist= d(j,D)$ \tcp*{ $d(.)$ returns distance between $j$ and $D$}
    	\If{$Dist \leq p_x$}
            {
                Insert $j$ in $S$
            }
    }
$L_r = RandomPicker(S)$ \tcp*{ randomly picks one from $S$}
return $L_r$\;

 \caption{Greedy approach for recommendation}
 \label{algo:greedy}
\end{algorithm}

\fi

Since \textit{MoParkeR} employs a non-dominated sort, it generates a set of trade-off recommendations considering four utilities associated with a parking lot: fare, walking to a destination, travel time, and parking likelihood. It is guaranteed to always pick parking lots with the lowest travel time, smallest walking distance to destination, lowest parking fare, and high likelihood in addition to other trade-off recommendations. On the other hand, the greedy algorithm rarely picks parking lots with these finest constraints as it employs a random approach. We run both approaches 100 times and log the proportion of times a parking lot with these finest constraints were chosen. \textit{MoParkeR} always selects parking lots with the smallest fare, travel time, walking, and the largest likelihood in both Melbourne and Rye datasets. In contrast, the greedy approach rarely picks parking lots with the lowest fare in Melbourne and Rye datasets (i.e. 14\% and 18\% cases respectively). For the smallest travel time, these ratios are 11\% and 14\% for Melbourne and Rye respectively. The greedy approach selects parking lots that require the smallest walking to destination in 13\% and 16\% cases in these two datasets. However, the likelihood of getting parking spaces available within the recommended solutions is very low (i.e. 7\% and 10\% in Melbourne and Rye datasets respectively).

\iffalse
\begin{figure}[h]
  \centering
  \includegraphics[width=.9\columnwidth]{./figures/Moparker-vs-greedy2}
  \caption{Proportion of finest utility selection}
  \label{fig:moparker-vs-greedy}
\end{figure}
\fi

\iffalse
\begin{figure}[h]
  \centering
  \includegraphics[width=\columnwidth]{./figures/Likelihood-prediction}
  \caption{Prediction of parking likelihood}
  \label{fig:Likelihood-prediction}
\end{figure}
\fi

We also show that the parking likelihood used for recommendation generation is predictable. The prediction API resides in the `Response generator' module of \textit{MoParkeR} implements three state-of-the-art time-series prediction models to both of our datasets. Table \ref{tab:prediction-results} shows that LightGBM outperforms the other two deep learning models for both datasets and ConvLSTM does a better job in the Melbourne data in terms of MAE. All methods achieve good results in parking availability prediction, which suggests that the parking vacancy is predictable. Therefore, seeking for a vacant parking slot in advance considering the likelihood is possible by using time-series prediction approaches.

% \usepackage{multirow}
\begin{table}[h]
\centering
\caption{ Prediction of parking likelihood}
\label{tab:prediction-results}
\scalebox{0.75}{
\begin{tabular}{lllllllll}
\toprule
\multicolumn{1}{c}{\multirow{3}{*}{\textbf{\begin{tabular}[c]{@{}c@{}}Prediction\\ Model\end{tabular}}}} & \multicolumn{4}{c}{\textbf{Melbourne data}}                                                                          & \multicolumn{4}{c}{\textbf{Rye data}}                                                                         \\ \cmidrule(lr){2-5} \cmidrule(lr){6-9} 
\multicolumn{1}{c}{}                                                                                     & \multicolumn{2}{c}{MAE}                                  & \multicolumn{2}{c}{RMSE}                                 & \multicolumn{2}{c}{MAE}                                  & \multicolumn{2}{c}{RMSE}                                 \\ \cmidrule(lr){2-3} \cmidrule(lr){4-5} \cmidrule(lr){6-7} \cmidrule(lr){8-9} 
\multicolumn{1}{c}{}                                                                                     & \multicolumn{1}{c}{15 min} & \multicolumn{1}{c}{30 min} & \multicolumn{1}{c}{15 min} & \multicolumn{1}{c}{30 min} & \multicolumn{1}{c}{15 min} & \multicolumn{1}{c}{30 min} & \multicolumn{1}{c}{15 min} & \multicolumn{1}{c}{30 min} \\ \midrule \midrule
LightGBM                                                                                                   & 0.0924                      & 0.1381                      & \textbf{0.1563}             & \textbf{0.2090}             & \textbf{0.0174}             & \textbf{0.0228}             & \textbf{0.0294}             & \textbf{0.0335}             \\ 
ConvLSTM                                                                                                   & \textbf{0.0882}             & \textbf{0.1231}             & 0.1667                      & 0.2268                      & 0.0368                      & 0.0324                      & 0.0430                      & 0.0426                      \\ 
LSTM                                                                                                       & 0.0889                      & 0.1253                      & 0.1640                      & 0.2282                      & 0.0355                      & 0.0321                      & 0.0420                      & 0.0422                      \\ \bottomrule
\end{tabular}}
\end{table}

%Parking likelihood prediction plays a crucial role in parking slot recommendation since we need to sense the vacant parking slot during a certain time.

\section{Discussion}
The primary goals of the paper are to formalize a novel problem called multi-objective parking recommendation and to present a solution by designing a multi-objective parking recommendation engine that considers various conflicting goals together. These goals have been achieved by formalizing a novel problem and employing a non-dominated sorting technique to calculate a set of recommended parking spots in the Pareto-front. %Several cities in Australia have been publishing their parking data for use by researchers. Specifically, we used data from the cities of Melbourne and Rye, all in Victoria, Australia. The reason why we selected these two cities for our study because of data availability and the number of car parks. In addition, these cities are selected because the city of Melbourne is considered a very busy business district while the city of Rye is a popular tourist spot. Thus, people can use our novel engine to decide where to go and where to park. 
The results show that our \emph{MoParkeR} is able to provide a wide range of reasonably good parking lots in terms of travel time, walking distance to the destination, parking fare, and parking likelihood. The developed \emph{MoParkeR} can make several implications and benefits to the users including:
\begin{itemize}
\item Improved total journey information provided by \emph{MoParkeR} could be ultimately accessible for users to find the optimal parking location will only assist to support the economic, environmental, and amenity of towns and cities. %Reduced congestion, vehicle emissions, and time spent seeking a parking bay are the immediate benefits that will be realised through more efficient trips. 

%\item %As highlighted in the previous study, the customers place high importance on accessibility and availability of parking facilities \cite{shoppingAus}. Businesses will also gain from greater customer generation with the availability of personalised user access information for parking which will result in better utilisation of available parking. 

\item Tourists and visitors will also be able to plan trips more effectively dependant on their individual needs for a more enjoyable visit to a high demand attraction consequently ensuring economic stimulus to regions.

\item Social benefits are also feasible with special users with for example that need to locate the convenient disabled parking bay that minimises walking distances and total trip time.

\item For parking management and planning,  \emph{MoParkeR} could be utilised to assess the ratings of various trip scenarios to establish the optimal parking fees and parking control time limits to match user tolerances and requirements.
    
\end{itemize}

%These benefits have a direct impact on social, time and economic perspectives. A crucial benefit would be that the user will require a less time in finding parking space which has a direct impact on economic and pollution. Consequently, the users will be healthier as they breathe clean air, less stress and having a happier life.  

%\sr{TODO: Implication/cost-benefits analysis for smart cities}
\section{Conclusion}

We presented a solution to the multi-objective parking recommendation problem. We adopted an epsilon-domination approach to compute trade-off parking lots in the Pareto front considering four objectives: total travel time, walking distance to destination, parking fare, and parking likelihood. We further refine the recommendations using a crowding distance and objective-threshold computation. The applicability of our \emph{MoParkeR} engine is illustrated by the deployment in two cities in Australia. Future research may include a user study to investigate the user experience of using such a multi-objective parking recommendation engine along with the inclusion of more objectives and ways to provide a reduced number of parking recommendations by considering the actions of other drivers. Parking allocation upon the recommendation in real-time is another avenue of research that will be inspired by our work.

\begin{acks}

This work is supported by the Australian Government and Mornington Peninsula Shire Council through the provision of a Smart Cities and Suburbs Grant.
\end{acks}

\bibliographystyle{ACM-Reference-Format}
\bibliography{sample-base}

\end{document}